\documentstyle[epsfig]{apj}
\include{epsf}
\title[\centerline{ Gravitational lens magnification}]
{\vglue-3.0truecm
\centerline{\it\small For publication in The Astrophysical Journal}
\vglue 2.5truecm
\noindent
\centerline{\Large GRAVITATIONAL LENS MAGNIFICATION AND }
\centerline{\Large THE MASS OF ABELL 1689 }
\author[\centerline{A.N. Taylor et al.}]
 {\centerline{\large A.N. TAYLOR, \, S. DYE}\\ 
 \centerline{Department of Astronomy, University of Edinburgh, 
Blackford Hill, Edinburgh EH9 3HJ, UK; ant@roe.ac.uk, sd@roe.ac.uk} \\
	\hspace{1.cm}\\
 \centerline{\large T.J. BROADHURST, \,  N. BEN{\'I}TEZ}\\
 \centerline{Department of Astronomy, Campbell Hall, 
University of California, Berkeley, USA; tjb@wibble.berkeley.edu; benitezn@wibble.berkeley.edu} \\
	\hspace{1.cm}\\
 \centerline{ AND}\\
	\hspace{1.cm}\\
 \centerline{\large E. VAN KAMPEN}\\
 \centerline{Royal Observatory Edinburgh, Blackford Hill, Edinburgh EH9 3HJ,}\\
\centerline{Theoretical Astrophysics Center, Juliane Maries Vej 30, 
DK-2100 Copenhagen, Denmark; eelco@tacsg1.tac.dk }}}
  
\newcommand{\be}{\begin{equation}}
\newcommand{\ee}{\end{equation}}
\newcommand{\ba}{\begin{eqnarray}}
\newcommand{\ea}{\end{eqnarray}}

\newcommand{\rgl}{\rangle}
\newcommand{\lgl}{\langle}
\newcommand{\hMpc}{{h^{-1} \rm Mpc}}
\newcommand{\Mpc}{\rm Mpc}
\newcommand{\hMsol}{{h^{-1} \rm M_\odot}}
\newcommand{\kms}{\,{\rm kms}^{-1}}

\newcommand{\nn}{\nonumber \\}

\newcommand{\bkappa}{\bar \kappa }
\def\bib{\parskip=0pt\par\noindent\hangindent\parindent
    \parskip =2ex plus .5ex minus .1ex}

\begin{document}

\maketitle
 
\begin{abstract}
We present the first application of lens magnification to measure the absolute
mass of a galaxy cluster; Abell 1689.
The absolute mass of a galaxy cluster can be measured by the 
gravitational lens magnification of a background galaxy population by
the cluster gravitational potential.
The lensing signal is complicated by the intrinsic variation
in number counts due to galaxy clustering and shot--noise, and by additional
uncertainties in relating magnification to mass in the strong lensing regime. 
Clustering and shot--noise can be dealt with using maximum likelihood methods.
Local approximations can then be used to estimate the mass from magnification.
Alternatively if the lens is axially symmetric we show that the 
amplification equation can be solved nonlocally for the surface mass density 
and the tangential shear.

\hglue 1.0truecm
In this paper we present the first maps of the total mass 
distribution in Abell 1689, measured from the deficit of lensed 
red galaxies behind the cluster. Although noisier, these reproduce the 
main features of mass maps made using the shear distortion of background
galaxies, but have the correct normalisation, finally breaking the ``sheet--mass'' 
degeneracy that has plagued lensing methods based on shear.
 Averaging over annular bins centered on the peak
of the light distribution we derive the cluster mass profile in the 
inner $4'$ ($0.48 \hMpc$). These show a profile with a near isothermal surface
mass density
$\kappa\approx (0.5 \pm0.1) (\theta/1')^{-1}$ out to a radius of 
$2.4'$ ($0.28\hMpc$), followed by a sudden drop into noise. We find that 
the projected mass interior to $0.24 h^{-1}$Mpc is 
$M(<0.24 h^{-1}{\rm Mpc})=(0.50 \pm 0.09) \times 10^{15} \hMsol$.

\hglue 1.0truecm
We compare our results with masses estimated from X-ray temperatures and
line-of-sight velocity dispersions, as well as weak shear and 
lensing arclets. We find that the 
masses inferred from X-ray,  line-of-sight velocity dispersions, 
arclets and weak shear
are all in fair agreement for Abell 1698.

\hfill\break

\end{abstract}


\section{INTRODUCTION}

The magnitude and distribution of matter in galaxy clusters should 
in principle provide a strong constraint on cosmological models 
of structure formation and the mean mass density of the Universe.
In addition a direct image of the mass density
 will tell us much about the relationship
between gas, galaxies and dark matter and whether light is indeed a 
fair -- if biased -- tracer of mass.

	Early techniques for estimating the mass in clusters 
include dynamical methods, from the line--of--sight
 velocity dispersion of member galaxies, and
X-ray temperature measurements. However these estimates make some
strong assumptions about equilibrium conditions in the cluster.

Kaiser and Squires (1993) circumvented this by showing that a more 
direct method of estimating
the mass, with no underlying assumptions about the dynamical or 
thermodynamical state of the cluster, was to measure the shear field
in the source distribution of the cluster  background (Kaiser
\& Squires 1993; Tyson, Valdes \& Wenk 1990, Schneider \& Seitz 1995 ). 
On average the shear pattern of a population of unlensed galaxies 
 should be randomly distributed. But in the
presence of a massive gravitational lensing cluster, the shear 
field is polarized. Since the shear field is related (non-locally) 
to the surface mass density the shear can be used to estimate the 
mass distribution -- up to an arbitrary constant. 
The presence of this arbitrary constant, referred to as the ``sheet--mass''
degeneracy, means that only differential masses can be measured. Shear maps
are conventionally normalised to the edge of the observed field, or such that 
the inferred mass--density is everywhere positive,  and so represent 
a lower limit on the mass.

Soon after, Broadhurst, Taylor \& Peacock (1995; hereafter BTP) showed 
that the sheet--mass degeneracy could be broken by use of the gravitational
lens magnification effect. The number and magnitude--redshift distribution of 
background galaxies is distorted by the gravitational field of the lensing cluster
and in the weak lensing regime this 
distortion provides a straightforward estimate of the surface mass density.
With calibration from off-set fields the cluster mass distribution can be 
properly normalized. 

BTP also suggested that a degraded, but much quicker, estimate
of the magnification effect could be made from the distortion of angular 
number counts of background sources. Broadhurst (1995) found evidence
for this distortion in the background counts of the cluster Abell 1689,
as did Fort, Mellier \& Dantel-Fort (1997) for Cl0024.
In this work we apply the methods developed by BTP and extended by 
Taylor \& Dye (1998) to estimate the surface mass density from the 
distortion of angular counts, including the effects of shot--noise and galaxy 
clustering, and van Kampen (1998) on estimating the   
surface mass density in the strong lensing regime, to Abell 1689.

The layout of the paper is as follows. In Section 2
we describe the magnification effect itself. In Section 3 we describe the 
effects of shot noise and clustering on estimates of the surface mass density.
In Section 4 we describe how to estimate the surface mass density
in the strong lensing regime using a local approximations, and introduce a
new, self--consistent, nonlocal solution for axially symmetric lenses.
We apply these methods to map out the mass in the cluster Abell 1689 in Section 5
and find its profile. Our mass estimate is compared to other estimates
in Section 6 and our conclusions are presented in Section 7.

\section{The Magnification Effect}
The observed number of galaxies seen in projection
on the sky is (BTP, Taylor \& Dye 1998)
\be
	n' = n_0 A^{\beta-1} (1+ \Theta ),
\label{eq:lensmag}
\ee
where $n_0$ is the expected mean number of galaxies in a given 
area at a given magnitude. Variations in this mean arise from the 
angular perturbation in galaxy density, $\Theta$, due to galaxy 
clustering, and from gravitational lens magnification. The lens
amplification factor is
\be
	A=|(1-\kappa)^2 - \gamma^2|^{-1}
\label{eq:ampequ}
\ee
where 
\be
	\kappa = \frac{\Sigma}{\Sigma_{\rm crit}}
\ee
is the surface mass density in units of the critical surface mass,
$\Sigma_{\rm crit}$. The amplitude of the shear field is given by $\gamma$ 
and the background galaxy luminosity function is locally approximated
by 
\be
n(L) \sim L^{-\beta}.
\ee
The amplification index, $\beta-1$, accounts for the 
expansion of the background image and
for the increase in number as faint sources are lensed above the
flux limit.

In the absence of galaxy clustering and finite sampling effects the 
background galaxy distribution can simply be inverted, 
via equation (\ref{eq:lensmag}), to find the amplification.
One can then solve equation (\ref{eq:ampequ})
to find the surface density, with some realistic assumptions about the shear. In 
Section 4 we discuss various approximations that allow us to do this.

However, given a small resolution scale for the surface amplification,
galaxy clustering and finite sampling will in general be an important effect. In the 
next Section we discuss the effects of intrinsic variation in
the distribution of the background galaxy sources.

\section{Galaxy Clustering Noise}

The main sources of uncertainty in lens magnification
 are due to shot-noise, from finite sampling, and the intrinsic clustering of
the background source population which introduce correlated
fluctuations in the angular counts. As we are viewing
small--angles the clustering properties of the background source 
galaxies are not in general linear, unless the depth of background
is sufficient to wash out the clustering pattern. As a result it is not
sufficient to make the usual assumption that galaxy clustering
can be modelled by a Gaussian distribution.

We can account for the effects of shot--noise and nonlinear clustering
by modeling the angular counts by a Lognormal--Poisson 
model (Coles \& Jones 1991, BTP, Taylor \& Dye 1998) -- a random 
point--process sampling of a lognormal density field. The distribution 
function of source counts is then
\ba
	P(n) &=& \frac{1}{n !} \lgl \lambda^n e^{-\lambda} \rgl, \\
	     &=& \frac{\lambda_0^n}{n !} \int^{\infty}_{-\infty} 
	\frac{dx}{\sqrt{2 \pi}\sigma} \,
		\exp \left( - \frac{x^2}{2\sigma^2} - \lambda_0 e^x -n x
		 \right)
\label{eq:LNdist}
\ea
where $\lambda= \lambda_0 e^x$  is the local mean 
density, $x$ is a Gaussian 
random variable of zero mean and variance $\sigma^2$ and 
$\lambda_0=n_0 A^{\beta-1} e^{-\sigma^2/2}$ correctly normalises the counts. 
The linear clustering variance, $\sigma^2$, is related to the nonlinear variance 
by $\sigma^2= \ln[1+\sigma_{\rm nl}^2]$. We have tested this distribution against 
available data and find that it is an excellent fit to the distribution of 
counts in the deep fields. The only parameters are the observed 
count per pixel, $n$,
and the variance of the lognormal field. The amplitude of clustering of
the density field and its dependence of redshift can be estimated from, 
e.g., the I-band selected galaxies
in the Canada--France Redshift Survey in the range $17.5<I<22.5$
(Le F{\'e}vre 1996; see section \ref{sectclust}). The quantity required 
is the variance in a given area of sky,
which can be estimated by averaging the observed 
angular correlation function, $\omega(\theta)$, over a given area:
\be
	\sigma_{\rm nl}^2 = \bar{\omega}(\theta)= \frac{1}{\Omega(\theta)} 
	\int_\Omega\! d^2 \theta' \, \omega(\theta'),
\label{angavw}
\ee
where $\Omega(\theta)$ is the area.

Our method of approach is then that discussed by BTP. At each 
pixel in a map of the source counts one uses the distribution 
equ. (\ref{eq:LNdist}) as a likelihood function, 
${\cal L}(A|n,\sigma)=P(n|\sigma,A)$, assuming a uniform prior for
the amplification. The surface density is then found from the amplification
by making some
realistic assumption about the shear and maximizing the likelihood. In the 
next Section we discuss a number of ways of transforming from the 
amplification to $\kappa$ in the strong lensing regime.

\section{The Strong Lensing Regime}

Transforming from amplification to the surface mass density is
potentially non-trivial, as we have no shear information. One could
incorporate this from independent measurements of the shear field, but 
for the present discussion we are interested in developing a completely 
independent lensing approach. We shall discuss combining shear and 
magnification elsewhere. In principle one could generate a first guess
for the surface mass density and iterate the amplification equation towards a 
solution of both surface density and shear. However, given the small
field of view and uncertainties introduced by parity changes, this can
be an unstable problem. In addition, as the solutions are in general
multivalued, we would hope to start from as near to the correct solution
as possible. In this section we discuss
a number of reasonable approximations to solve the amplification
equation (\ref{eq:ampequ}). These can be regarded as solutions in 
their own right, or as the first best guess to an iterated solution.
We begin by discussing the local approximation methods suggested and 
fully tested on simulated clusters by van Kampen (1998). Then in Section 4.2
we present a new, self--consistent solution to the amplification equation, for 
$\kappa$ and $\gamma$ for an axially symmetric lens.

\subsection{Local approximations to the surface mass density}

There exist only two local relations between $\gamma$ and $\kappa$ that
result in a single caustic solution of the amplification equation (2)
which is easily invertible (van Kampen 1998): $\gamma=0$, corresponding
to a sheet of matter, and $\gamma=\kappa$, for an isotropic lens.
These two relations have corresponding estimators for $\kappa$ as a function
of amplification:
\ba
\label{eq:approx2}
   \kappa_0\ \equiv\ \kappa(\gamma=0) &=& 1 - {\cal P}A^{-1/2},  \\
\label{eq:approx1}
   \kappa_1\ \equiv\ \kappa(\gamma=\kappa) &=& \frac{1}{2} (1-{\cal P}A^{-1}),
\ea
where ${\cal P}=\pm1$ is the image parity.

Let us assume that the surface mass density of the lens is smooth 
over some scale. In this 
case, for a sufficiently smooth lens $\gamma \leq \kappa$ (BTP). The 
equality holds in the case of an isotropic lens, for
instance the isothermal lens. The inequality holds for any anisotropic
lens, with the sheet mass at the extreme. For a smooth lens these two
estimates bound the true value 
	$\kappa_1 \leq \kappa \leq \kappa_0$. Before caustic
crossing it can also be shown that 
	$\kappa_1 \leq \kappa_0 \leq \kappa_{\rm weak}$ holds, where 
$A = 1+2\kappa_{\rm weak} $ is the weak lensing limit (BTP). 
Hence the weak lensing approximation will overestimate the cluster
mass in the strong regime, usually by a factor of two (van Kampen 1998).

In practice, substructure and asphericity of the cluster will induce extra
shear (e.g.\ Bartelmann, Steinmetz \& Weiss 1995), especially in the
surrounding low-$\kappa$ neighbourhood, where substructure is relatively
more dominant and filaments make the cluster most aspherical. This means
that the lens will not be smooth for small $\kappa$, and therefore $\kappa_1$
is a lower limit for the true $\kappa$ only for the central parts of the
cluster, in the case the lens parity is known. Van Kampen (1998) found it to
be a good lower limit only for $\kappa>0.4$ (for the most massive clusters),
while for $\kappa<0.2$, $\kappa_1$ is usually fairly close to the true value.
For angle--averaged $\kappa$-profiles $\kappa_1$ is a good lower limit for
$\langle\kappa\rangle_\theta > 0.2$.
All this has no bearing on $\kappa_0$, which remains a strong upper limit
until the first caustic crossing.

An approximation that tries to take these cluster lens features into
account, while still giving an invertible $A(\kappa)$ relation, is
(van Kampen 1998):
\be
	\gamma = |1-c| \sqrt{\frac{\kappa}{c}},
\ee
which results in an amplification relation that admits the full
four solutions:
\be
	A^{-1} = |(\kappa-c)(\kappa-1/c)|,
\ee
with caustics at $\kappa=c$ and $1/c$. The solution for $\kappa$ is then
\be
	\kappa_c = \frac{1}{2c}\left((c^2+1)-{\cal S} 
	\sqrt{(c^2+1)^2-4c^2(1-{\cal P}A^{-1})}\right).
\label{eq:twocaustic}
\ee
Solutions are set by choosing the
parities ${\cal P,S}=\pm 1$ where ${\cal P}$ is the image parity
and ${\cal S}$ is the sign of $((c^2+1)/2c-\kappa)$.
Note that the sheet--like solution is recovered by setting $c=1$.

\begin{figure}[h]
\epsfxsize=9.cm
\epsfysize=8.cm
\vspace{-0.cm}
\hfill
\epsfbox{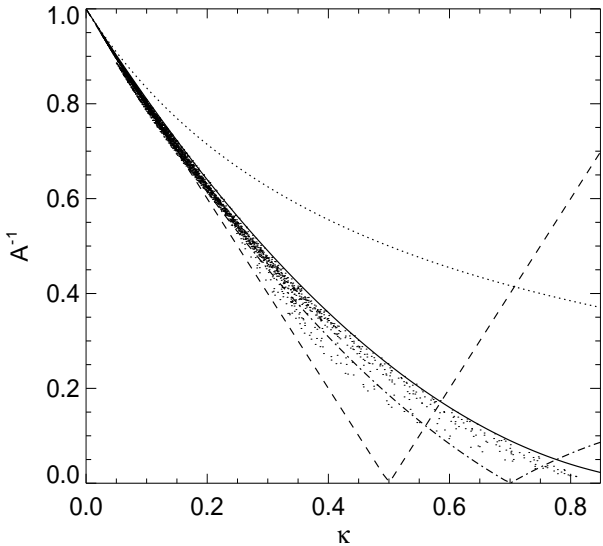}
\hfill
\epsfverbosetrue
\vspace{0.cm}
\small\caption{Scatter plot of the surface mass density, $\kappa$, versus
the inverse amplification, $A^{-1}$, for a simulated cluster in a CDM
universe (see van Kampen 1998 for details). The cluster is at a redshift
of $0.183$ and the background population at $z=0.8$. The cluster was
selected to look similar to Abell 1689. The solid line is the $\gamma=0$ strong
lensing  approximation. Before caustic crossing this is a hard bound on
the locus of points. The dashed line is the $\gamma=\kappa$ approximation,
which is a good lower bound for $\kappa>0.2$ for this cluster.
The weak--lensing approximation (dotted--line) is seen to be a very
bad approximation for $\kappa>0.1$.
The dot-dashed line is a best fit to the simulation for the 2-caustic 
approximation $\gamma =|1-c|\sqrt{\kappa/c}$, with $c=0.7$.}
\label{fig1}
\end{figure}

Figure 1 shows a plot of $\kappa$ versus the inverse amplification,
$A^{-1}$, for the three estimators. Also shown is the weak field approximation.
The points are taken from a simulated
lensing cluster (van Kampen \& Kargert 1997) which is of comparable size
to A1689. It is clear that $\kappa_0$ is a strong bound, at least
until a caustic is crossed, and that $\kappa_1$ provides a very good bound
for $\kappa>0.2$. The weak field approximation however is extremely bad
except in the very weak regime ($\kappa<0.1$). The two--caustic approximation
behaves as it is designed to do: a good fit between the other two strong
lensing estimators for the central parts of the cluster, while also
modelling the $\gamma>\kappa$ behaviour for small $\kappa$. These
results are fairly robust over a wide range of clusters and for all
realistic values of the cosmological density parameter $\Omega_0$.

\subsection{A nonlocal approximation to the surface mass density}

An alternative  approach is to assume axial symmetry for the lens.
Because this fixes a non-local functional relationship between 
$\kappa$ and $\gamma$ (equation \ref{eq:gamma-kappa}), we can  
solve the amplification equation (\ref{eq:ampequ}) for a 
self--consistent  $\kappa$ and $\gamma$ profile.  
Although we shall apply our results to circularly averaged data, these results
hold for any self-similar embedded set of contours.

We define a mean surface density interior to a contour by
integration over the interior area, $\Omega(\theta)$,
\be
	\bkappa(\theta) = \frac{1}{\Omega(\theta)} 
		\int_{\Omega} d^2 \theta' \,\kappa(\theta') 
\ee
The deflection angle for the axi-symmetric lens is 
\be
	\Delta \theta = \theta \bkappa,
\ee
and the shear is given by
\be
	\gamma = \gamma_t =| \kappa - \bkappa |.
\label{eq:gamma-kappa}
\ee 
where the tangential term, $\gamma_t$, is the only component of
shear generated. 
The amplification factor is given by
\be
	A^{-1} = |(1- \bkappa)(1-2 \kappa + \bkappa)|.
\ee

One can now simultaneously solve for the 
surface mass density, shear and amplification by series
solution. Firstly we divide the surface mass into consecutive
shells with equal separation (any arbitrary separation can be used, we
have chosen a regular separation for convenience).
If we split $\bkappa$ into an interior term, $\eta_{n-1}$,
and a surface term, then for the $n^{th}$ shell we have
\be
	\bkappa_n = \eta_{n-1} + \frac{2}{(n+1)}\kappa_n
\ee
where we have defined
\be
	 \eta_{n-1}=\frac{2}{n(n+1)} \sum_{m=1}^{n-1}m\kappa_m,
\ee
The surface mass density in the $n^{th}$ shell is then given by
\ba
	\kappa_n &=& \frac{(n+1)}{4n}\big(n+1-(n-1)\eta_{n-1}- \nn 
	 & & {\cal S} [
	(n-1-(n+1)\eta_{n-1})^2+4n{\cal P} A_n^{-1}]^{1/2}\big),
\label{axikappa}
\ea
where ${\cal P}$, ${\cal S}=\pm 1$ are again the image parities.
The only freedom we have, for a given amplification profile is
the choice of the shear on the first shell, $\gamma_1=\eta_0$, and the parity.
It should be noted that given the amplification and having fixed the parities 
one has to ensure that the first $\gamma$ satisfies 
$\gamma^2 \geq {\cal P} A^{-1}$, to avoid unphysical solutions.
The nonlocal approximation contains both the sheet and isothermal 
solutions as specific solutions.
 The uncertainty on $\kappa$ and $\gamma$ can be found
by simple error propagation of the uncertainty on the
measurement of the amplification.

Having shown in Sections 2, 3 and 4 how, in principle, one can measure 
the surface mass density from angular number counts, in the next 
section we exploit these methods to measure the mass distribution 
in the lensing cluster Abell 1689.

\section{Application to Abell 1689}

In this section we apply the methods discussed in sections 2, 3 and 4
to observational data. We begin by describing the data.

\subsection{The Data}

\subsubsection{Data acquisition and reduction}
The data were obtained during a run in February 1994 at ESO's NTT 3.6m telescope,
with $10^4$ secs integration in the V and I bands, and 
covers $70$ square arcminutes on the cluster. Seeing was similar in
both bands, with FWHM of $0.8''$ and a CCD pixel scale of $0.34''$.
The EMMI instrument was used throughout. The passbands and exposures were
chosen such that the cluster E/S0 galaxies would be bluer than a good
fraction of the background, requiring much deeper imaging in the bluer 
passband for detection. The cluster was observed down to a limiting 
magnitude of $I=24$.

The images were debiased and flattened with skyflats using 
standard IRAF procedures. After this there remained some 
large scale gradients of a few percent, probably caused by some 
rotation of the internal lens. We additionally corrected each 
separate exposure with a smoothed version of itself, obtained 
after masking out the cluster and other bright objects. Following
this we had homogeneous photometry across the field (A further
discussion of the reduction procedure can be found in 
Ben\'\i tez, et al, 1998, in preparation).
The zero point was found to be good to $0.1$ magnitudes. High 
humidity on a few nights meant that some of the data was not 
photometric, so we calibrated with the photometric data. 
The object detection and classification was performed with SExtractor.

\subsubsection{Separation of cluster and background}
\begin{figure}
\centering
\begin{picture}(200,200)
\includegraphics{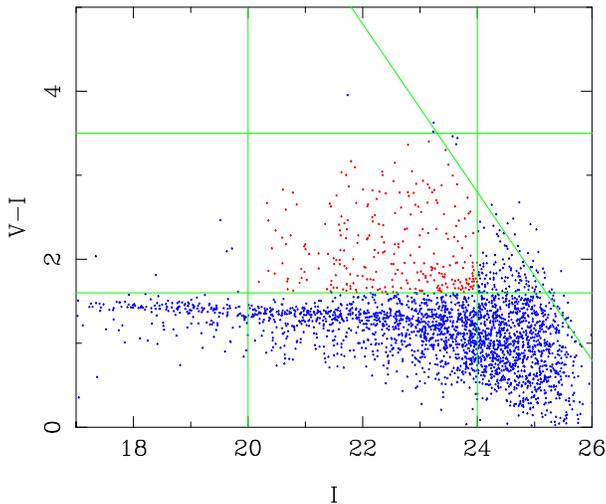}
\end{picture}
\caption[]{Colour--magnitude diagram for A1689, overlaid with 
colour cuts used to isolate the cluster members from the background
population; $20<I<24$, $1.6<V-I<3.5$ and $V<26.8$. 
The strong horizontal band of galaxies is the cluster E/SO sequence.}
\label{colormag}
\end{figure}

To measure the distortion in background counts we must first 
separate the background
from cluster members and mask off the area they obscure.
Cluster galaxies were identified from the strong cluster E/SO 
colour sequence, which forms a horizontal band across the colour--magnitude
diagram, shown in figure \ref{colormag}. The sharp upper edge of this band 
represents the reddest galaxies in the cluster. Galaxies redder than this 
are cosmologically redshifted, and hence represent a background population. 
As well as isolating cluster members, this selection should also ensure 
that any foreground galaxies are removed.
Anything redder than $V-I=1.6$ was selected as a background 
galaxy. Further colour cuts where imposed to ensure completeness of
the sample. The range of magnitudes was restricted to $20<I<24$ and 
the $V$-band limited to $V<28$. Finally we also cut at $V-I<3.5$, where 
the reddest galaxies cut off.

Since the identification of cluster members is important to remove
contamination of the background sample we also checked our colour 
selected candidates with new data from a photometric redshift survey 
of the same field (Dye et al, 1998, in preparation). We found general 
agreement with the simpler colour selection.

\begin{figure}
\centering
\begin{picture}(200,200)
\includegraphics{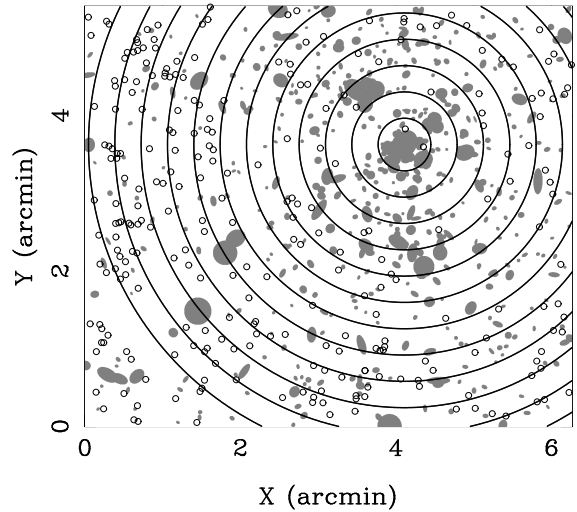}
\end{picture}
\small\caption{The masked region of A1689. Cluster members where selected
using colour information (see text) and then  masked over so that these regions
do not affect the surface density estimate of background sources. The
total region masked is about $10\%$ of the area. The 
background galaxies are also shown as open circles. Superimposed are the 
concentric bins used to calculate the radial profile, centered on the
peak in the light distribution. North is up and east is to the left.}
\label{figmask}
\end{figure}

Having identified foreground and cluster members we produced a mask to eliminate 
those areas obscured by cluster members that would otherwise bias the 
mass estimate. To isolate
the cluster members for the mask we selected all the galaxies in the colour--magnitude
diagram lower than $V-I=1.6$ and less than $I=22$. This isolated most of the 
cluster sequence. The remaining galaxies in the region 
$V-I<1.6$ and $I>22$, $V<26.8$,
we identified as the faint blue background population. 
It is clear from figure \ref{colormag} that the distinction between faint
cluster member and faint blue background galaxy is rather vague. However, since
the faint cluster members are also the smallest, the masked area is fairly
insensitive to the exact division.
Figure \ref{figmask} shows the distribution of cluster galaxies and the red background
population. The concentric circles are centered on the peak in the cluster 
light distribution and show the position of the annuli used to calculate the 
radial profile in section 5.4.

\subsubsection{Selection by colour}

Once the cluster galaxies have been isolated,
the background galaxies may be sub-divided into a red and blue population, 
separated by $V-I=1.6$. The observed slope of the luminosity function for 
these two populations for $I>20$ is $\beta_R=0.38$ and $\beta_B=1$ 
(Broadhurst 1995; we shall do a more accurate fit using our colour cuts
in section 5.2.2). From equation 
(\ref{eq:lensmag}) we expect that the surface density of red galaxies will be 
suppressed due to the dilation effect, while magnification of the faint 
blue galaxy population will compensate for the dilation. 
Hence selecting by colour allows us to identify a population 
of galaxies with a very flat luminosity function to boost the
lensing signal, at the expense of a reduction in galaxy numbers.
Simple error analysis shows that the signal--to--noise 
varies as (Taylor \& Dye 1998)
\be
	S/N = 2 |\beta-1| \kappa A (1-\kappa+\gamma'/\kappa') 
	\sqrt{n} (1+n\sigma^2)^{-1/2},
\label{eq:s/n}
\ee
where $'\equiv\partial/\partial R$.
While the signal--to--noise is a linear function of the slope
of the luminosity function, it only grows with the square--root of 
the galaxy numbers, assuming Poisson statistics. Hence one can get a better 
signal-to-noise by pre-selection of the red background population to boost 
the signal, at the expense of numbers.
Equation (\ref{eq:s/n}) also shows that one can get a better signal
by observing to fainter magnitudes to simultaneously enhance the surface 
number density and reduce the contribution from intrinsic clustering 
(see Taylor \& Dye, 1998, for a more detailed discussion of observing strategies). 

There is also a practical reason for favouring the red galaxy population.
While the cluster members are unlikely to be redder than the cluster E/SO
sequence the distinction between faint blue galaxies and cluster members, based
on selection from the colour--magnitude diagram alone, is somewhat vague.
There may be blue cluster members that will contaminate
the sample of blue background galaxies. In the absence of redshift information
the blue background population is clearly harder to isolate.

As we have noted the red population has relatively few faint counts,
so that the expansion term in equation (1) dominates and there is a net
underdensity of red galaxies behind the cluster (see Fig. \ref{redbackground}
and Fig. \ref{fig4}). 
Conversely, faint blue galaxies are numerous and cancel the expansion 
effect. As expected we found the blue galaxies were uniform across the A1689
field. This is a good indicator that it is the magnification effect at work, 
and not some spurious contaminant, for example colour gradients across
the field, or large scale variations due to clustering. In addition it 
also indicates that the deficit in the red population is not due to dust 
obscuration or reddening in the cluster, as this would affect both red 
and blue populations in equal measure.

\subsection{The distribution of background galaxies}

\begin{figure}
\centering
\begin{picture}(200,200)
\includegraphics{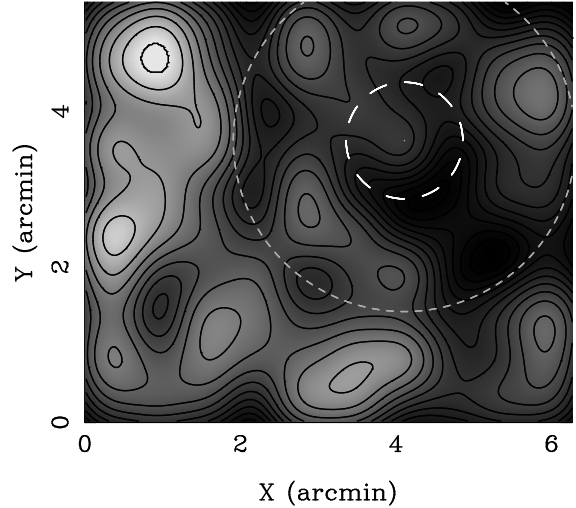}
\end{picture}
\small\caption{The distribution of red I-band background sources for
Abell 1689. Darker grey areas indicate an underdensity of source counts.
The image is Gaussian smoothed with a smoothing scale of $0.35'$. 
The peak of the light distribution is at $(4.1',3.6')$. The maximum
density of objects is $23.0$ per square arcminute and the minimum is 
$1.1$ per square arcminute. There are 15 contour lines spaced by $\Delta n =1.46$
galaxies per  square arcminute. A 
strong caustic feature is seen $0.75'$ from the peak (inner dashed line), 
more visible to the south--west, as the other side of the peak is masked over.
A second feature is found in the radial profile at $2.2'$ (outer dotted line).
The image is orientated with East to the left and North to the top.}
\label{redbackground}
\end{figure}

  In Figure \ref{redbackground} we show the surface distribution of the red
population behind A1689, Gaussian smoothed on a scale of $0.35'$. There
are 268 background galaxies. The cluster
members have been masked out and the masked areas interpolated over.
The masked region contributes to only $\approx 10\%$ of the total field. 
Figure \ref{figmask} shows the masked region.
The cluster center, identified as the peak of the light distribution,
is at $(4.1',3.6')$.

 The angular size of the cluster scales like 
\be
	R(\theta)=0.87 D_A(z_c) (\theta/1') \hMpc,
\ee
where $D_A(z)=2(1-(1+z)^{-1/2})/(1+z)$ is the comoving, dimensionless 
angular distance in an Einstein--de-Sitter
universe. Hence at the redshift of Abell 1689, 
 $z_c=0.183\pm0.001$ (Teague et al 1990), one arcminute is about
$0.117 \hMpc$.

Figure  \ref{redbackground} clearly shows a deficit of galaxies 
about the central peak in the light distribution at $(4.1',3.6')$. At
$\theta = 0.75'$ there is an arc of very underdense number counts
to the South--West of the cluster center, marked by a dashed line (The
background is somewhat obscured by the cluster mask to the North--East of
the cluster center). This 
is clear indication of a caustic feature in the background number counts,
where the number density drops to zero due to dilation. This exactly
corresponds to the radius of the blue arcs observed by Tyson \& Fisher (1995)
at $\theta=0.85'$ (see also the radial number counts in section 5.4). This
is  strong evidence that we have detected the magnification effect in the
background counts.

\subsubsection{The redshift distribution of background galaxies}

The efficiency of lensing varies with the redshift of the background
source (BTP). Therefore it is important to estimate the background
redshift distribution.
Crampton et al (1995) find that Canada--France Redshift Survey (CFRS)
has a  median redshift of $z=0.56$ for galaxies in the range $17.5<I<22.5$. 
They also show a colour--redshift 
diagram that indicates that the red galaxy population ($V-I>1.6$) has a 
median redshift of about $z\approx0.8$ (Crampton et al 1995, their figure 5). 
More accurately we can integrate the best fit Schechter function 
found by Lilly et al. (1995) for the CFRS red galaxy population. This has 
parameters $\phi^*=0.0031\pm0.00095$,
$\alpha=1.03\pm0.15$ and $M^*(B)=-21$, where 
$M(B)=I-5\log_{10}(D_L/10{\rm pc})+2.5 \log_{10}(1+z)+$k-correction, where
the k-correction is discussed in their paper and $D_L=(1+z)r(z)$ is the luminosity
distance. Lilly et al. found no detectable evolution 
of the luminosity function of the CFRS red population and we assume no evolution. 
Extrapolating to the magnitude range $20<I<24$ we find that the redshift distribution 
can be well fitted by the function 
\be
		\phi(z) = \frac{\alpha z^2}{z_*^3 \Gamma[3/\alpha]}
			\exp(-(z/z_*)^\alpha)
\ee
with $\alpha=1.8$ and $z_*=0.78$ to  about $5\%$ accuracy over a redshift 
range of $0.25<z<1.5$. The moments of this distribution are 
\be
	\lgl z^n \rgl = z_*^n \frac{\Gamma[(3+n)/\alpha]}{\Gamma[3/\alpha]}.
\ee
Hence for the red galaxy population we find that $\lgl z\rgl=0.96$ and 
$\sigma_z=0.42$. To simplify the analysis of the lensing properties of the cluster,
we shall assume hereafter that the background distribution is at a single
redshift of $z=0.8$ (the mode of the distribution)
and has an uncertainty of $\delta z = 0.4$.

As the caustic indicated by the blue arcs coincides with the magnification 
caustic, we can presume that the galaxy forming the arcs lies at the same 
redshift as the magnified red background galaxies, $z \approx 0.8$. At 
present we do not know the redshift of this arc.

\subsubsection{Number counts of the background galaxy population}

\begin{figure}
\centering
\begin{picture}(200,200)
\includegraphics{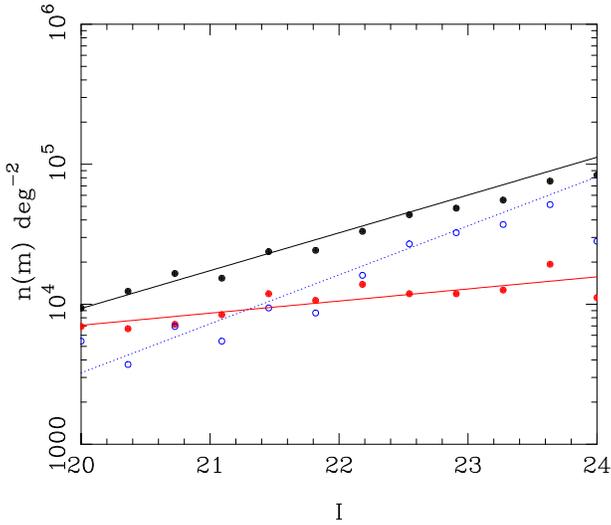}
\end{picture}
\caption[]{Magnitude distribution of all I-band galaxies (solid dots),
the red selected galaxies (grey dots) and the blue background galaxies 
(open dots). The lines are the best fits to the
data.}
\label{figkeck_counts}
\end{figure}

Of major importance to the lens magnification method is the normalisation of the
background galaxy population. The CFRS is not adequate for this, since their 
colour cuts were in the rest frame $U-V$, rather than the observed $V-I$.
Instead we have used the Keck data of Smail et al (1995), who observed deep
$VRI$ images down to a limiting magnitude of $R\approx27$. The total differential
galaxy count rate in the I-band can be approximated by
\be
	\log_{10} n = (0.271 \pm 0.009) I  -  1.45
\ee
over the range $20<I<24$, where $n$ is per magnitude per square degree. We have 
applied our colour criteria (see section 5.1) to the Keck data 
and find that the red galaxy population $V-I>1.6$, can be well approximated 
by
\be
		\log_{10} n({\rm red}) = (0.0864\pm0.0187)I +  (2.12 \pm 0.41)
\ee
over the range $20<I<24$. Figure \ref{figkeck_counts} shows the magnitude
distribution for the full dataset and for the red--selected galaxy population
and the best--fit lines. Integrating the fit for the red galaxies 
yields a total count rate of $n=12.02\pm3.37$ 
galaxies per square arcminute in the range $20<I<24$. Since 
$\beta=2.5\, {\rm d} \log_{10} n /{\rm d} m$, we find that the Keck data implies 
$\beta_R=0.216\pm0.047$. This is the value of $\beta$ we shall use in the
subsequent analysis. 

An alternative, although less exact, method of
normalisation is to assume negligible cluster mass at the edge of the field
and normalise the cluster to this. In general this would put a lower
limit on the mass, and is similar to the method used to normalise shear
mass maps. In fact
if we do this for A1689 we find a background count rate very similar
to that given by the Keck data. The error introduced into the final
mass estimate by uncertainties in $\beta$ scale as 
$\delta \kappa/\kappa \approx\delta \beta/|1-\beta|$, which for the 
Keck data results in a fractional error of around $5\%$.

We have also fitted the blue counts in the Keck sample 
(Figure \ref{figkeck_counts}). Over the same
range as the red counts we find that 
$\log_{10} n({\rm blue}) \approx 0.35 I -3.49$,
resulting in $\beta_B = 0.88$, close to the lens invariant $\beta=1$,
 and a count density between $23<I<24$ of $n_0({\rm blue})=15.5$ galaxies per 
square arcminute.

\subsubsection{Clustering properties of the background population}
\label{sectclust}

The amplitude of clustering of I-band galaxies and its dependence on 
redshift can be estimated from
the  CFRS (Le F{\'e}vre et al. 1996).  Le F{\'e}vre (1996) find that 
there is little difference between the clustering properties of 
red and blue populations of galaxies for $z>0.5$, implying that the 
populations were well mixed at this epoch. We therefore apply their
clustering results directly to our red galaxy population. They fit their
results to a power--law model for the evolving correlation function,
$\xi(r)=(r/r_0)^{-\gamma}$, where
\be	
		r_0(z) = r_0(0) (1+z)^{-(3+\varepsilon)/\gamma}
\ee
where $\varepsilon=1\pm1$ and $r_0(z=0.53)=1.33\pm0.09 \hMpc$ 
and $\gamma=1.64\pm0.05$
is in this section the slope of the correlation function.

The quantity we require is the variance in a 
given area of sky, which can be estimated by averaging the observed 
angular correlation function, $\omega(\theta)$, over a given area
(equation \ref{angavw}).
The clustering variance for I-band galaxies then scales roughly 
like (Taylor \& Dye 1998) 
\be
	\sigma_{\rm nl}^2 = 10^{-2} z^{-2.8} (\theta/1')^{-0.8}
\ee
where the sampled area is a circle of radius $\theta$ and
we have assumed unbiased, linear evolution of the density 
field. The background galaxies are assumed to all lie at $z \approx 1$.

\subsection{Reconstructing the surface mass density}

\begin{figure}[h]
\centering
\begin{picture}(200,200)
\includegraphics{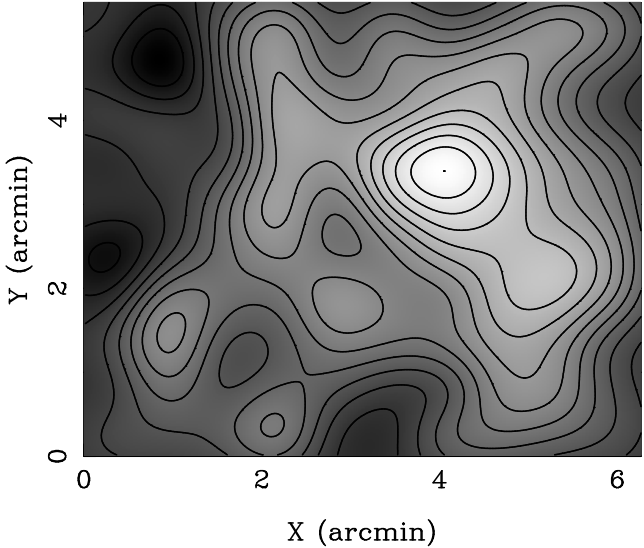}
\end{picture}
\caption[]{Reconstruction of the surface mass density of Abell 1689 from
the red background galaxy population,
using the nonlinear local sheet approximation and a full likelihood
analysis in 2-D. Light regions are high density.  Only one caustic line is 
assumed, at $\theta=0.75'$ from the peak of the light distribution.
The maximum surface density is $\kappa=1.35$, at ($4.02',3.41'$), consistent
with the peak in the light distribution. The minimum surface mass
density is $\kappa=-0.47$.
There are 15 linearly spaced contours, separated by $\Delta \kappa = 0.12$ and 
the map is Gaussian smoothed with a smoothing length of $\theta_S=0.35'$.
North is up and east is to the left.}
\label{fig3}
\end{figure}

In Figure \ref{fig3} we plot the reconstructed surface mass density of Abell 1689
using the nonlinear local sheet approximation, $\kappa_0$ (see section 4.1), 
changing parity on the caustic line at $\theta=0.75'$
(see Figure \ref{redbackground}). 
The uncertainty on the peak of the mass distribution
is somewhat large (see Section 5.2), but significant features can be
seen around the cluster core. There appears to be an extension
to the south--west not seen in the cluster galaxy distribution. Interestingly
there also appears to be a loosely connected ridge, about $2.4'$ from the peak.
We shall discuss this feature further below, but note that the shear mass 
map derived by Kaiser (1996; figure 2) shows similar extensions and 
ridge, although the extension to the west is not apparent in the shear map.
Two underdense regions are also seen to the south  and to the 
east in both maps. While the comparison is only qualitative and the 
maps are noisy, we find this very encouraging as these maps are derived 
from completely independent methods, although the underlying data set is the same.

\subsection{The Mass Profile of Abell 1689}

\begin{figure}[h]
\centering
\begin{picture}(200,200)
\includegraphics{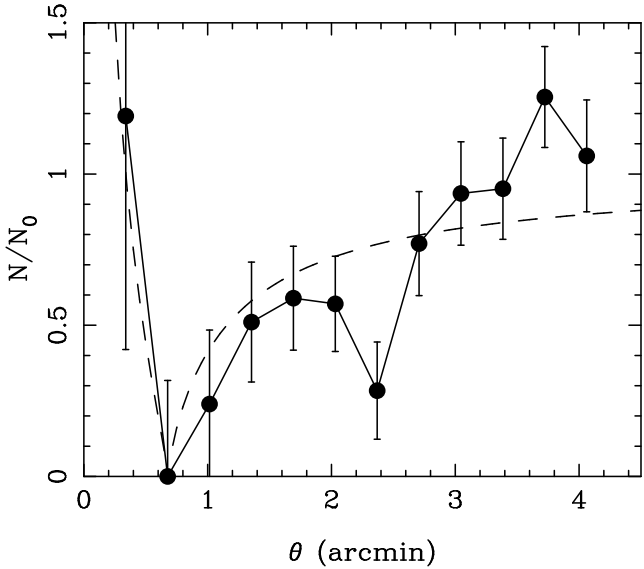}
\end{picture}
\small\caption{Radial profile of red counts behind Abell 1689. The background
count density is  $n_0= 12$ objects/arcmin$^2$. Superimposed is the profile 
for an isothermal model, normalised at the caustic radius, $\theta=0.75'$
(dashed line).}
\label{fig4}
\end{figure}

\begin{table}
\begin{tabular}{|l|l|l|l|l|} 	\hline
$r$(arcmin)&    N   &   N/N0 &Annulus area & Obscured area  \\ \hline
  0.33     &    2   &   1.19 &      0.35   &   0.21   \\  
  0.67     &    0   &   0.00 &      1.08   &   0.25   \\  
   1.01    &    4   &   0.24 &      1.79   &   0.40   \\  
   1.35    &   13   &   0.51 &      2.51   &   0.39   \\  
   1.69    &   20   &   0.59 &      3.23   &   0.40   \\  
   2.03    &   23   &   0.57 &      3.62   &   0.26   \\  
   2.36    &   11   &   0.28 &      3.46   &   0.23   \\  
   2.70    &   26   &   0.77 &      3.08   &   0.26   \\  
   3.04    &   32   &   0.94 &      3.01   &   0.16   \\  
   3.38    &   34   &   0.95 &      3.14   &   0.17   \\  
   3.72    &   45   &   1.25 &      3.18   &   0.20   \\  
   4.06    &   31   &   1.06 &      2.49   &   0.06 \\ \hline
\end{tabular}
\small\caption{Table of angular radius ($r$ in arcminutes), number of red galaxies 
($N$), ratio of galaxies to background ($N/N_0$), the total area of the annuli  
(areas are in arcmin$^2$) and area obscured by the mask. The unobscured
area is total area $-$ obscured area. The expected number of galaxies in an 
annuli is  $N_0 = n_0 \times$ unobscured area}
\end{table}

While the mass maps are suggestive, a more quantitative measure can be 
made by angle averaging the counts and calculating the mass profile.
Figure \ref{fig4} shows the radial counts about the 
peak in the light distribution, normalised to the Keck data.
The plotted error bars are only due to Poisson statistics, although in the mass
analysis below we shall take into account the effects of clustering. 
A general trend is clear, and lies close to the prediction for an isothermal
lens normalised to the blue arc caustic. This has a surface mass
density of $\kappa=0.375 (\theta/1')^{-1}$, corresponding to a virial
velocity of $1600\kms$. Again it is worth emphasizing that the zero of
the number counts at $\theta=0.75'$ corresponds to the caustic inferred
from the blue arcs. The second dip will be discussed in more detail 
in section 5.4.4. The increase in counts at $\theta=3.7'$ is likely
to be due to a clustering effect.

\begin{figure}[h]
\centering
\begin{picture}(200,200)
\includegraphics{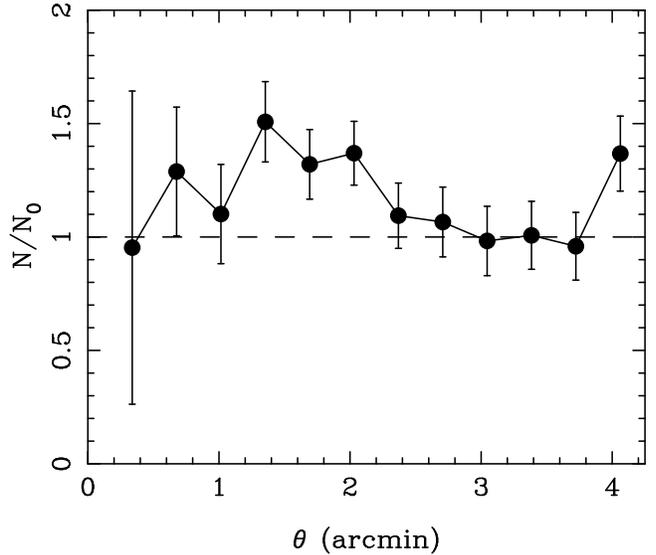}
\end{picture}
\small\caption{Radial profile of blue counts behind Abell 1689. The background
count density is  $n_0= 22$ objects/arcmin$^2$. As expected from the nearly
lens invariant slope $\beta_{\rm blue}=0.88$, the number counts are 
nearly flat, and at large radii tend towards $n/n_0=1$. The slight increase
towards the cluster center is probably due to contamination of the counts
by blue cluster members}
\label{figbluecounts}
\end{figure}

In Figure \ref{figbluecounts} we show that radial profile for the blue 
galaxy population, normalised with the Keck data in section 5.2.2. 
As expected there is no lensing signal. The slight increase towards 
the cluster center is due to contamination from the blue cluster
members.

\begin{figure}
\centering
\begin{picture}(200,200)
\includegraphics{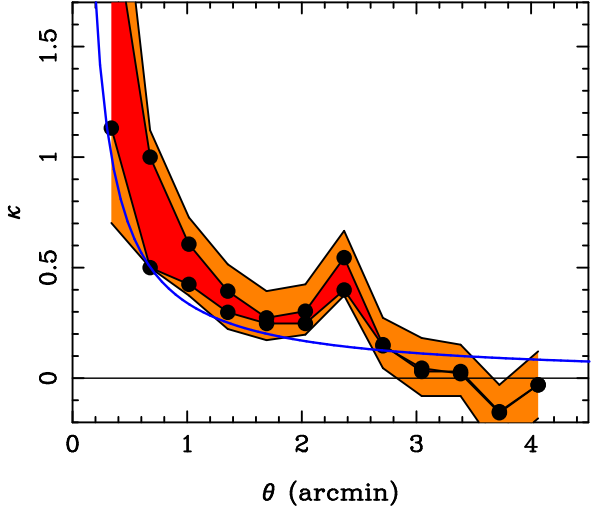}
\end{picture}
  \small\caption{Radial profile of surface mass density of 
cluster Abell 1689. The dark solid region shows the uncertainty
due to the strong lensing estimators. The lighter shaded region 
is due to the clustering and shot-noise uncertainty of the
background population. The solid line is a singular isothermal
profile, normalised to the caustic feature at $\theta=0.75'$. }
\label{fig5}
\end{figure}

\subsubsection{Local approximations for the surface mass density}

Fig. \ref{fig5} shows the radial mass profile of the cluster Abell 1689
assuming a single caustic at $\theta=0.75'$.
The two solid lines are calculated using the Lognormal--Poisson likelihood 
estimator (equation \ref{eq:LNdist}) with each of the two strong lensing 
approximations (equations \ref{eq:approx1} and \ref{eq:approx2} ). The light
shaded region indicates the $1\sigma$ uncertainty due to both
shot noise and the effects of clustering. The dark shaded region
indicates the region between the two extreme estimators. Away from the
cluster center these agree and are equal to the weak lensing 
estimator, but noise effects become dominant. Closer to the cluster 
center the uncertainty due to the shear increases and becomes
dominant at $\theta < 1'$. However the cluster mass profile is significantly
detected between $1'<\theta < 2.6'$. We also appear to see a deviation from an
isothermal profile, which is also plotted.
When the procedure was repeated with the center of the annuli offset 
from the peak of the light distribution the mass profile was 
weaker and less significant, as one would expect if the peak
of the mass density was associated with that of light.

\begin{figure}
\centering
\begin{picture}(200,200)
\includegraphics{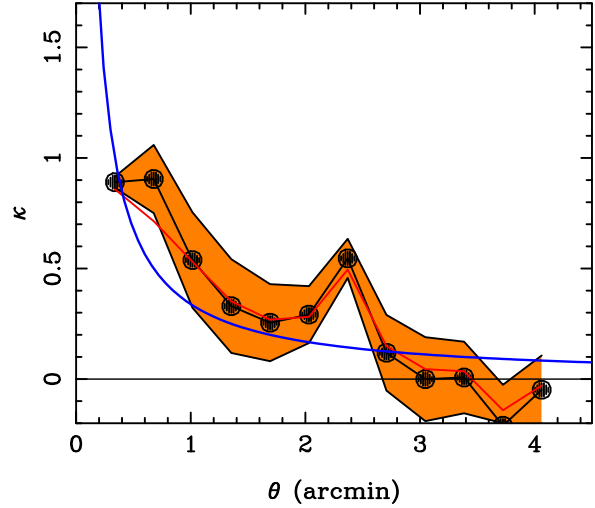}
\end{picture}
  \small\caption{Radial profiles of  surface mass density, $\kappa$,
  for A1689 (solid line with dots), calculated by solving the 
 axially symmetric lens equation (\ref{axikappa}). The shaded regions are $1\sigma$
 errors calculated via error propagation from the uncertainty on the
 measured amplification profile. The solid dark line is a singular isothermal 
 profile normalised to the caustic feature at $\theta=0.75'$. The 
 lighter solid line is the local best fit estimator, $\kappa_c$}
\label{fig6}
\end{figure}

\begin{figure}
\centering
\begin{picture}(200,200)
\includegraphics{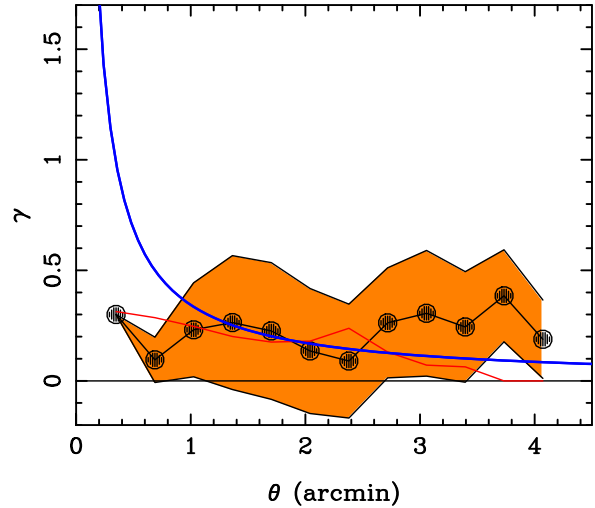}
\end{picture}
  \small\caption{Radial profiles of 
 tangential shear, $\gamma_t$, for A1689 (solid line with dots), 
 calculated by solving the 
 axially symmetric lens equation (\ref{axikappa}). The shaded regions are $1\sigma$
 errors calculated via error propagation from the uncertainty on the
 measured amplification profile. The solid dark line is a singular isothermal 
 profile normalised to the caustic feature at $\theta=0.75'$. The 
 lighter solid line is the local best fit estimator, $\kappa_c$.}
\label{figgamma}
\end{figure}

\subsubsection{Nonlocal approximation for the mass density and shear}

In Figures \ref{fig6} and \ref{figgamma} we assume axisymmetry and 
equation (\ref{axikappa}) to calculate
the surface mass--density and shear simultaneously. We set
$\gamma_1=0.3$ for the first shell. The resulting profile is fairly
insensitive to this choice, only affecting the first 2 shells. The
uncertainty on the shear in the first shell is small because
this must be chosen a priori. However averaging over shells means that
the errors do not strongly propagate through to higher radii. Again a
mass detection is found between $1'$ and $2.8'$, this time with 
the shear accounted for. In this region $\kappa \approx 0.4 \pm 0.15$,
which is somewhat higher than that found by the shear estimate of
$\kappa =0.2\pm0.1$ (Kaiser 1996; note that we quote Kaiser's colour selected 
sample, where cluster members that may contaminate the shear estimate 
have been removed. This corresponds to combining our red and blue background
populations. This will change the redshift distribution of the background
and include some residual blue cluster contamination which may account for the 
discrepancy). Also, for the single caustic solution,
we see a large spike at $2.2'$, which is not seen in the Kaiser (1996)
results. However the shear method correlates points, which may lead
to both the suppression of features, and underestimation of the errors.

Our estimate of the shear field is far more uncertain, with $\gamma_t=0.2\pm0.3$
over most of the range. There is a slight increase beyond $2.4'$ due
to the spike in the surface mass profile at that radius, but the profile 
is dominated by noise. 
This increase is not reflected in the angle averaged
measurements of Kaiser (1996), where the mean shear is $\gamma=0.15\pm0.05$.

\subsubsection{Local approximation for the surface mass density and shear}

Figures \ref{fig6} and \ref{figgamma} also show $\kappa$ and $\gamma$
estimated from the best-fit solution of section 4.1. We find good agreement
 between the local and nonlocal approximations for $\kappa$, but the the 
shear profiles 
are somewhat different, reflecting that one estimator is local and one
nonlocal. However, the large uncertainties produced by each estimator
means that we cannot predict the shear profile with much certainty
from the available data.

\subsubsection{Two background populations ?}

An interesting feature of the counts in Figure \ref{fig4} is the appearance of
two pronounced dips, one at $0.75'$ and another one at $2.2'$. 
While the inner dip has already been identified with a  
caustic line, the outer dip is somewhat anomalous. A number of 
possibilities could account for this. The feature was noted in the 
mass plot as a low signal-to-noise ridge in the density and can be 
seen in the number counts as a loosely connected ring about the cluster 
center. One possibility is that 
this is due to clustering in the background population, combined with 
a large mass concentration to the south--east of the peak in the 
light distribution. There are few cluster members in the region
of the ridge, or the bump, so the effect is not due to masking.

An alternative is that this is the first glimpse of a second caustic line. 
In principle a second caustic can be created by placing the background 
galaxies at two redshifts,
one at low redshift, one at high redshift (eg, Fort, Mellier \& Dantel-Fort 1996). 
 The observed number counts would then be given by
\be
	n'/n_0 = A^{\beta-1}_1 + 
	\nu\big(A^{\beta-1}_2 -A^{\beta-1}_1\big) 
\ee
where $A_i=A(f_i)$ with
$f_i=\kappa(z_i)/\kappa_\infty=(\sqrt{1+z_i}-\sqrt{1+z_L})/(\sqrt{1+z_i}-1)$
(BTP), and $i=1,2$ for the two galaxy populations. $\nu$ is the
fraction of galaxies at redshift $z_2$. An outer caustic line must 
be produced by the high redshift population. If we make this population
lie at $z=0.8$, then the low redshift population must lie at $z=0.3$. 
Both populations are reflecting the same arc, the difference in 
projected radii is
wholly due to their relative redshifts. 

However this would double the predicted mass from lens magnification, making
Abell 1689 a very extreme cluster. In addition it
seems hard to make a caustic line from the high redshift population
for such a massive cluster without forming a second, inner radial caustic.
As the strongest arc is tangential and is seen near the inner arc, one
would have to conspire to have a nearby galaxy, at $z=0.3$, lensed
and lying at the same projected radii as the radial arc produced 
by the high redshift population. This seems highly unlikely.

	One could also keep the mass roughly constant and place a second population
at $z>0.8$. This is a possibility, but does not strongly affect our mass
estimate assuming a single caustic solution. In the absence of further evidence
for a second high redshift population, we shall only consider the single
caustic model.

\subsection{Mass estimate of Abell 1689}

\subsubsection{From $\kappa$ to mass surface density}

Assuming that the background galaxies all lie at the same redshift of
$z=0.8$, and given that the surface density scales as 
\be
	\Sigma = \frac{1}{3} \Sigma_0 \kappa 
	\frac{(\sqrt{1+z}-1)(1+z_L)^2}
	{(\sqrt{1+z_L}-1)(\sqrt{1+z}-\sqrt{1+z_L})},
\ee
where $\Sigma_0=8.32 \times 10^{14} h M_\odot \Mpc^{-2}$ is 
the mean mass per unit area in the universe, then we find that
the surface mass density is 
\be
\Sigma=5.9 \times 10^{15} \kappa \, [h M_\odot \Mpc^{-2}].
\ee
Although we have assumed an Einstein--de-Sitter universe,
these results only depend weakly on cosmology (BTP).

\subsubsection{Uncertainty in the redshift distribution}

The error introduced by assuming the background galaxies lie at the
same redshift can be estimated by error propagation and assuming
$\delta z=0.4$ (see section 5.2.1). Hence 
$\delta \Sigma = |\partial \Sigma/\partial z| \delta z$ and the fractional
uncertainty on the surface mass density due to the uncertainty in redshift
distribution of the background galaxies is 
$\delta \Sigma/\Sigma = 0.37 \delta z = 0.148$. The same error is also 
found in mass estimates based on the shear pattern.

\subsubsection{Uncertainty due to normalisation of background counts}

Assuming a sheet mass solution ($\kappa_0$ in section 4.1) we find that 
the uncertainty due to the normalisation of the background counts is
$\delta \kappa = (|1-\kappa|/2|\beta-1|) \delta n_0/n_0$. For A1689
and the red galaxy population this is $\delta \kappa = 0.15 |1-\kappa|$.
For an average $\kappa=0.5$ the uncertainty is around $\delta \kappa=0.07$.

\begin{figure}
\centering
\begin{picture}(200,200)
\includegraphics{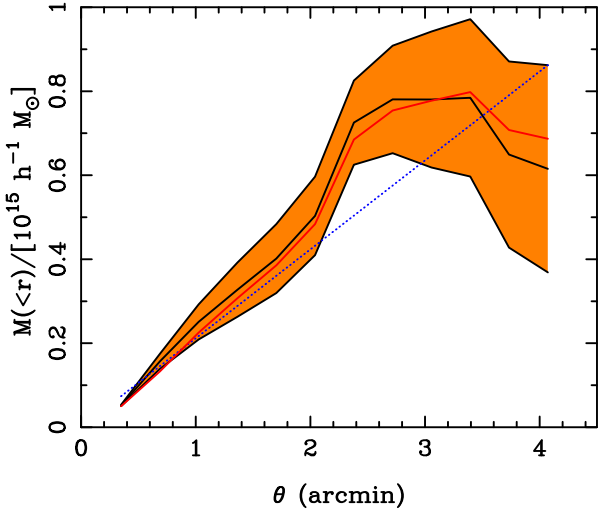}
\end{picture}
  \small\caption{Cumulative mass profile of Abell 1689. The solid dark line
and shaded uncertainties are estimated using the axi-symmetric nonlocal
estimator described in section 4.2. The lighter grey line is the cumulative mass
estimated from the best--fit local approximation, $\kappa_c$ described in section 4.1. Also plotted is the isothermal fit to the blue arc caustic (dotted line), 
similar to the shear results of Kaiser (1996) and Tyson \& Fischer (1995).}
\label{fig7}
\end{figure}

\subsubsection{The cumulative mass distribution}

Figure \ref{fig7} shows the cumulative mass interior to a shell,
calculated from both the nonlocal approximation (section 4.2), and
the best--fit local approximation allowing only a single caustic solution
(section 4.1). 
The uncertainties are treated by error propagation.
We find that the 2d projected mass interior to $0.24 \hMpc$ is  
\be
	M_{2d}(<0.24\hMpc)=(0.50 \pm 0.09) \times 10^{15} \hMsol,
\ee
 and that
the two estimators are in good agreement. We find that the projected mass
scales like 
\be
	M_{2d}(<R)\approx 3.5\times 10^{15} (R/\hMpc)^{1.3}\hMsol,
\label{magprofile}
\ee
for $R<0.32 \hMpc$, 
similar to that for an isothermal sphere; $M \sim R$. Hence it appears that 
A1689 has a near--isothermal core. Beyond $R=0.32 \hMpc$ the lensing
signal is lost in background noise, and we can only say that $\kappa \leq 0.1$

Including the uncertainty from the background redshift distribution 
and the normalisation of background counts increases the error to about $30\%$.

\section{Comparison with other mass estimates of A1689 and inferring
the 3d mass distribution}

In this section we compare the mass derived from lens magnification 
with that found from a number of other, independent measurements. 
Firstly we compare our results with estimates of the mass based on 
the shear pattern found around A1689 (section 6.1). The magnification and shear
complement each other in that the shear pattern has a higher signal--to--noise,
since it is not affected by clustering noise (although with redshift
information the magnification can also be measured free from clustering
noise -- see BTP), but suffers from the ``sheet--mass'' degeneracy.
We shall combine the magnification and shear pattern elsewhere.

While the lens magnification mass is vital for fixing the total 2d projected
mass distribution independently of any assumptions about the dynamical 
state of the cluster, much information can be gained by combining this
with other mass estimates, assuming that these are not strongly biased
by their reliance on thermodynamical equilibrium. 
In this section we describe a method for transforming from the 2d lens
mass to other cluster characteristics, such as the line--of--sight 
velocity dispersion (section 6.3) and the X-ray temperature
(section 6.4). Discrepancies that arise between these predicted 
characteristics and the actual measurements can be used to infer 
information about the mass distribution along the line of sight 
(Bartelmann \& Kolatt, 1997). We find that while 
there is fair agreement between all of the mass estimates when projection 
effects are taken into account, the agreement is better if the 
cluster A1689 is composed of two clusters superimposed along the line of sight
and separated by about $\Delta z=0.02$.

The transformation from a 2-d projected lensing mass to a 3-d mass,
line--of--sight velocity dispersion and X-ray temperature can be made using 
either the isothermal model, or by using relations 
found in N-body simulations of clusters. While the former provides a 
simpler method, one has more freedom with simulations to include or exclude 
the various projection effects that contaminate measurements of these quantities.
In this section we shall use the relations found by van Kampen (in preparation) 
from an ensemble of CDM cluster simulations, all with $\Omega_0=1$ and 
$\sigma_8=0.54$. These relations are model dependent, but serve to 
aid comparison between the various mass measurements.
We have also provided a table of quantities (Table 1) where the 
uncertainties have been calculated by combining the error on 
the cluster mass with the dispersion found in the deprojection relations.

We begin by comparing the lens magnification mass with the mass determined
from the shear field around A1689.

\subsection{Comparison with arclets and weak shear}

Tyson \& Fischer (1995) provide mass profiles of A1689 from arclets, another
independent estimator of the mass, normalized to the caustic line 
indicated by the blue arcs. They find that the 2d projected mass within 
$R=0.1 \hMpc$ is
\be
	M_{2d}(<0.1)=(0.18 \pm 0.01) \times 10^{15} \hMsol.
\ee
They also find that the mass scales like an isothermal sphere out 
 to $0.4 \hMpc$, before turning over to an $R^{-1.4}$ profile.
This implies that in the regime we probe with the magnification 
the cumulative mass scales like 
\be
	M(<R)=(1.8\pm0.1)\times 10^{15} (R/ \hMpc) \hMsol.
\ee
This is very close to the profile we find from lens magnification
(equation \ref{magprofile}). 
Using this we scale their results giving 
\be
	M_{2d}(<0.24)=(0.43 \pm 0.02) \times 10^{15} \hMsol,
\ee
in good agreement with the mass from magnification.
 
Kaiser (1996) has also calculated $\kappa$ based on the weak 
shear method (Kaiser \& Squires, 1993), using the same data we have used
here for A1689. We noted above that there are qualitative similarities
between the weak shear maps and those presented by Kaiser, which is
significant since the methods are independent.  
The mass--density profile found from the shear pattern is 
also well fitted by an isothermal profile; 
\be
	M_{2d}(<R)=1.8 \times 10^{15} (R/\hMpc) \hMsol,
\ee
with a $10\%$ statistical uncertainty and further $10\%$ systematic 
error due to the uncertainty in the redshift distribution (section 5.5.2).
Compared with our 2-d mass Kaiser's analysis suggests that 
\be
	M_{2d}(<0.28)= (0.43 \pm 0.04) \times 10^{15} \hMsol,
\ee
again in good agreement with that found by the magnification method.

\subsection{The 3d mass estimated from lensing alone}

The 3-d mass inferred from the 2-d projected mass inside a sphere 
of radius $r=0.5 \hMpc$ is 
\be
	M_{3d}(<0.5)=(0.72 \pm 0.25) \times 10^{15}\hMsol,
\ee
while the mass inside an Abell radius, $r=1.5 \hMpc$, is
\be
	M_{3d}(<1.5)=(1.6 \pm 0.6) \times 10^{15}\hMsol.
\ee
These estimates are probably an overestimate of the true 3d mass since the
dispersion in the simulations includes the effect of the alignment of the
clusters' principle axis along the line of sight. Given that the inferred
3d mass is so high A1689 is probably lying at the extreme of such a distribution.
In such cases the 3d mass may be much lower than mass inferred from a 2d 
projection. We discuss this possibility in the next few sections.

\subsection{Velocity dispersion of Abell 1689}

The predicted line--of--sight velocity dispersion estimated from the
simulations also includes the effects of superposition of clusters, 
infall along filaments and interlopers and so predict larger velocities 
than isolated clusters would. Given this we find that the observed line--of--sight 
velocity dispersion inferred from the 2d lensing mass is
\be
	\sigma_v(<1.5\hMpc) = 3400 \pm 900 \kms
\ee
for A1689. Again this is very much on the high side for a cluster 
in a typical CDM universe, and much higher than observational data suggests.

The velocity dispersion measured in a similar manner by 
Teague et al (1990) is $2355^{+238}_{-183} \kms$, nearly $1\sigma$ lower 
than our estimate. Hence the predicted velocity 
is not too different from the measured value, so long as both are 
measured in the same way. The large mass implied by lensing
suggests that A1689 is not a single cluster, but a superposition
of two smaller clumps. Miralde-Escud\'e \& Babul (1995) have
modeled A1689 by two adjacent clumps with velocity dispersions $1450\kms$ 
and $700\kms$, suggesting that one clump has a mass 4 times that of the
other. Taken together and including the relative velocity of the merging clumps,
this accounts for the larger velocities measured earlier.
den Hartog \& Katgert (1996) have tried to take into account interlopers 
and find that $v=1861 \kms$.

If, like Miralde-Escud\'e \& Babul, we assume a double cluster model
but place one cluster at $z=0.18$ with a velocity dispersion of
$1500\kms$ and the second at $z=0.20$ with a velocity dispersion of $750\kms$,
we find that we can reproduce a total projected velocity dispersion of
around $2300\kms$. Figures 4 and 5 of Teague et al (1990) also provide marginal 
evidence for a second concentration of galaxies at $z=0.2$. Hence it seems 
plausible that the high lensing mass of A1689 can be explained by a 
superposition of clusters.

\subsection{X--ray mass estimates of Abell 1689}

Evrard, Metzler \& Navarro (1996) have found that the mass within the 
radius defined where the mean cluster density is $500$ times the critical
density is strongly correlated with the cluster temperature. They fit 
this relation from simulations by
\be
	M_{500} = 1.11 \times 10^{15} \left( \frac{T_X}{10 {\rm keV}}\right)^{3/2}
        \hMsol,
\ee
where $T_X$ is the broad--beam temperature and $M_{500}$ is the 3d mass
within a radius defined by an overdensity $500\rho_{\rm crit}$. This 
radius is roughly given by $r_{500}=1.175 \hMpc$. 

X-ray temperatures of A1689 have been measured by both GINGA and ACSA.
Yamashita (1994) has analysed this data and finds $T=9\pm1$keV
while Mushotzky and Scharf (1997) find $T=9.02^{+0.4}_{-0.3}$keV. 
Daines et al (1997) have also recently re-analyzed ROSAT PSPC observations 
and find a mean temperature of $T_X=10.2 \pm 4 $keV. Note that we are quoting
the mean temperature, and incorporated the $40\%$ uncertainty
in the error estimate, rather than quoting upper limits as Daines et al do.
The major uncertainty in measuring X-ray temperatures here is instrumental,
as $10$keV is approaching the limit of ROSAT's sensitivity. 

Taking the result of Yamashita and the relation found by Evrard et al, 
we find that 
\be
	M_{500}=(0.95 \pm 0.16) \times 10^{15} \hMsol.
\ee
Using the simulated scaling relations we find 
\be
	M_{500}=(1.6 \pm 0.65) \times 10^{15} \hMsol
\ee
for Abell 1689, implying an X-ray temperature of $T_X=12.7\pm 3.4{\rm keV}$,
within the $1\sigma$ uncertainty of the measured X-ray 
temperature. Again, if we consider A1689 as a double cluster,  the 
nearer, larger mass concentration would be detected in X-ray, lowering 
the expected X-ray temperature. From the velocity dispersions we can infer a
temperature nearer to $T_X=0.7{\rm keV}$, slightly below, but again in agreement 
with observations.

\begin{table}
\begin{tabular}{|l|l|l|l|} 	\hline
Quantity        & This work        & Other 	   &		        \\ \hline
$M_{2d}(<0.24)$ & $0.50\pm0.09$    & $0.43\pm0.02$ &(Tyson \& Fischer)  \\ \hline
                &                  & $0.43\pm0.04 $& (Kaiser)           \\ \hline
$M_{3d}(<0.5)$  & $0.72 \pm 0.25$  &               &                    \\ \hline
$M_{500}$       & $1.6\pm0.65$     & $0.94\pm0.16$ & (Yamashita)        \\ \hline
$\sigma_v(<1.5)$& $3400\pm900$     & $2355^{+238}_{-183}$&(Teague et al) \\ \hline
\end{tabular}
\small\caption{Mass estimates for A1689 based on lens magnification 
(second column)
 and from other measurements (third column). 
Masses are given in units of $10^{15}\hMsol$ and velocities are quoted 
in units of $\kms$. Distance are given in $\hMpc$.
The other measurements are based
on arclets (Tyson \& Fischer 1995), the  shear pattern (Kaiser 1996), 
X-ray temperatures 
(Yamashita 1994) and line--of--sight
velocity dispersion (Teague et al 1990). Also given are the 3d mass
estimates from lens magnification. }
\end{table}

In conclusion, although we find a high mass,
there is a general consistency between the mass of A1689 
estimated from lens magnification and shear. In addition we
find a fair agreement between the lens mass and the line of sight
velocity dispersion if we take into account projection effects.
Modelling A1689 as a double cluster we find that the velocity
dispersion can be much lower, implying two smaller clusters, with the
lensing mass a superposition of cluster masses. This hypothesis might
also help explain the marginal discrepancy with X-ray temperature.

 However, if A1689 is a double
cluster, one would expect that the measured velocity dispersion would
be higher than that inferred by the lensing mass, due to the cluster 
separation, while the X-ray mass is lower, since only the more massive 
cluster dominates the X-ray emission. Our results indicate that Abell 1689
has a larger lensing mass than that implied by both velocity dispersion
and X-ray temperature, although given the large uncertainties it
is hard to be conclusive.

Finally, A1689 is in projection a highly spherical cluster, 
in contrast with the majority of clusters which appear extended. While this 
may be be a result of its high mass, it is also possible that A1689 
has its major axis aligned along the line of sight, pointing towards 
a second cluster. While much of the evidence on the mass distribution
along the line--of--sight is circumstantial, 
all of these effects would conspire to give A1689 its impressively massive 
appearance.

\section{Discussion}

The absolute surface mass density of a galaxy cluster can be estimated 
from the magnification effect on a background population of
galaxies, breaking the ``sheet--mass'' degeneracy. To apply this in 
practice, we have taken into account
the nonlinear clustering of the background population and shot--noise,
both of which contribute to uncertainties in the lensing signal. 
A further complication is the contribution of shear to the magnification
in the strong lensing regime, where the magnification signal is stronger. 
We have argued that this can be 
circumvented by approximate methods that can be either local,
where a relationship between surface mass and shear is assumed,
or by a nonlocal approximation where only the shape of the 
cluster is assumed. Both approximations seem to work well on simulated
data.

	We have applied these methods to the lensing cluster 
Abell 1689, using Keck data of Smail et al (1995) to normalise the 
background counts, and the CFRS results to infer the redshift distribution and 
clustering properties of our data. Using a $\gamma=0$ approximation  of the
surface density in the strong lensing regime, we have reconstructed a
2-d mass map for A1689 in the innermost 27 square arcminutes,  where a 
substantial part of the lensing signal comes from.
The 2d map has general features similar to those seen from shear 
maps (Kaiser 1996). This is encouraging for both methods, as they are 
independent determinations of the mass distribution.

For a more quantitative measure we have binned the data in annuli around the
peak in the light distribution and found a significant ($5\sigma$) drop
in the number counts, dropping to zero where a caustic is inferred from 
arcs. Local and nonlocal approximations were used to find the $\kappa$ profile 
from the number counts, and estimate the shear field. We found these
to be quantitatively similar to that found by the shear method.

 We have also
discussed the possibility of a second population of background galaxies,
creating a second dip in the radial number counts and a spike in the 
mass profile. However we argued that is is unlikely that there is a
second low-$z$ population as the cluster mass would be improbably high,
and if there is a high-$z$ population it has little effect on our results.

We have calculated a cumulative mass profile for A1689 and find a
projected 2-d cumulative mass of 
\be
	M_{2d}(<0.24\hMpc)=(0.50 \pm 0.09) \times 10^{15} \hMsol.
\ee
Such a large mass is very rare in a CDM universe normalised to the 
observed cluster abundance, and may indicate that A1689
is composed of two large masses along the line of sight, and/or
filaments connected to the cluster and aligned along the line--of--sight. 
This is  also implied by the high line of sight velocity dispersion 
which would be enhanced by merging clusters (Miralde-Escud\'e \& Babul 1995)
or by infall from aligned filaments.

	We have compared our mass estimates with other estimates
available in the literature and find  that the lens magnification,
shear, arclets, line--of--sight velocity dispersions and the X-ray 
temperature mass estimates are all in reasonable agreement, to within the 
uncertainties at this time.

The results presented here are from 3 hrs integration on the 3.6m NTT. 
Longer integration times have the combined benefit of increasing the number
of background galaxies, and so reducing shot noise, and of reducing 
the contribution from cosmic variance (equation \ref{eq:s/n}, Section 5.1.3).
Hence by increasing the exposure time we can expect to reduce the uncertainty
from lens magnification by a factor of 2 or so.

One drawback of this analysis is the contribution of clustering 
noise to the background counts. This can be removed using redshift
information, either from spectroscopy, or more efficiently using
photometric redshift information (BTP). We shall explore this elsewhere
(Dye et al, in preparation).

If our results are extended to other clusters we can hope to 
have a good representation of the total mass distribution, gas and
galaxy contents with which to make strong statistical arguments
about the matter content of the largest gravitationally collapsed 
structures in the universe.

\section*{Acknowledgments}

ANT thanks the PPARC for a research associateship and the 
University of Berkeley and the Theoretical Astrophysics Center,
Copenhagen for their hospitality during the writing of this paper. 
EvK acknowledges an European Community Research Fellowship as
part of the HCM programme and  thanks the University of Edinburgh 
and ROE for their hospitality. This work was supported in part by
Danmarks Grundforskningsfond through its funding of the Theoretical
Astrophysics Center.
SD thanks the PPARC for a studentship. NBL thanks the Spanish 
MEC for a Ph.D. scholarship, the University of Berkeley for 
their hospitality and financial support from the Spanish DGES, 
project PB95-0041.  We thank Ian Smail who kindly provided us
with a copy of his Keck data. 
We also thank John Peacock, Alan Heavens, Jens
Hjorth, Adrian Webster and an anonymous referee for comments
and useful suggestions.

\bigskip
\noindent{\bf REFERENCES}
\bib \strut

\bib Bartelmann M., Steinmetz M., Weiss A., 1995, AA, 297, 1
\bib Bartelmann M., Kolatt T.S., 1997, MNRAS, submitted, astro-ph/9706184
\bib Broadhurst T.J., Taylor A.N., Peacock J.A., 1995, ApJ, 438, 49
\bib Broadhurst T.J., 1995, in AIP Conf. Proc. 336, ``{\em Dark Matter}'',
 eds S.S. Holt, C.L. Bennett (New York)
\bib Coles P., Jones B., 1991, MNRAS, 248, 1 
\bib Crampton D., Le F\`erve O., Lilly S.J., Hammer F., 1995, ApJ, 455, 96
\bib Daines S., Jones C., Forman W., Tyson J.A., 1997, ApJ, submitted
\bib den Hartog R., Katgert P., 1996, MNRAS, 279, 349
\bib Evrard et al., 1996, ApJ, 469, 494
\bib Fort B., Mellier Y., Dantel-Fort M., 1997, AA, 321, 353
\bib Kaiser N., 1996, in ``{\em Gravitational Dynamics}'', Proc.
 of the 36thh Herstmonceux Conf., eds O. Lahav, E. Terlevich, R.J. Terlevich
(Cambridge University Press)
\bib Kaiser N., Squires G., 1993, ApJ, 404, 441
\bib Lilly S., Tresse L., Hammer F., Crampton D., Le F{\'e}vre O., 1995, ApJ,
455, 108
\bib Le F{\'e}vre O., Hudon D., Lilly S.J., Crampton D., Hammer F., Tresse L., 
1996, ApJ, 461, 534
\bib Miralda-Escud\'e J., Babul A., 1995, ApJ, 449, 18
\bib Mushotzky R.F., Scharf C.A., 1997, ApJ 482, L13
\bib Navarro, J.F., Frenk, C.S., White, S.D.M., 1996, ApJ, 462, 563
\bib Schneider, P. \& Seitz, C., 1995, A\&A, 294, 411
\bib Smail I., Hogg D.W., Yan L., Cohen J.G., 1995, ApJLett, 449,L105
\bib Taylor A.N., Dye S., 1998, submitted MNRAS
\bib Teague P.F., Carter D., Gray P.M., 1990, ApJSupp, 72, 715
\bib Tyson J.A., Valdes F., Wenk R.A., 1990, ApJ, 349, L1
\bib Tyson J.A., Fischer P., 1995, ApJLett, 446, L55 
\bib van Kampen E., 1998,  submitted MNRAS
\bib van Kampen E., Katgert P., 1997, MNRAS, 209, 327
\bib Yamashita, 1994, in Recontr. de Moriond on ``Clusters of Galaxies'',
eds F. Durret

\end{document}